# High-resolution intracellular recordings using a real-time computational model of the electrode

Romain Brette[*,1,2], Zuzanna Piwkowska[*,1], Michelle Rudolph-Lilith[1], Thierry Bal[1], Alain Destexhe[1]

* These authors contributed equally to this work.
1: Unité de Neurosciences Intégratives et Computationnelles (UNIC), CNRS, 91198 Gif-sur-Yvette, France
2: Équipe Odyssée (ENS/INRIA/ENPC), Département d'Informatique, École Normale Supérieure, F-75230 Paris Cedex 05, France

Correspondence should be addressed to Romain Brette brette@di.ens.fr.

Running title: High-resolution intracellular recordings

**SUMMARY**

Intracellular recordings of neuronal membrane potential are a central tool in neurophysiology. In many situations, especially *in vivo*, the traditional limitation of such recordings is the high electrode resistance, which may cause significant measurement errors. We introduce a computer-aided technique, Active Electrode Compensation (AEC), based on a digital model of the electrode interfaced in real time with the electrophysiological setup. The characteristics of this model are first estimated using white noise current injection. The electrode and membrane contribution are digitally separated, and the recording is then made by online subtraction of the electrode contribution. Tests comparing AEC to other techniques demonstrate that it yields recordings with improved accuracy. It enables high-frequency recordings in demanding conditions, such as injection of conductance noise in dynamic-clamp mode, not feasible with a single high resistance electrode until now. AEC should be particularly useful to characterize fast phenomena in neurons, *in vivo* and *in vitro*.

# INTRODUCTION

Intracellular recordings of neuronal membrane potential ($V_m$) are currently the only tool for studying the integration of excitatory synaptic inputs, inhibitory inputs, and intrinsic membrane currents underlying the spiking response. *In vivo*, as well as in some slice preparations, these recordings are done using either single high resistance sharp microelectrodes (Steriade et al., 2001; Crochet et al., 2006; Wilent and Contreras, 2005a; Higley and Contreras, 2007; Haider et al., 2007; Paz et al., 2007; Thomson and Deuchars, 1997; Wolfart et al., 2005) or single patch electrodes showing a high series resistance (Borg-Graham et al., 1998; Wehr and Zador, 2003; Mokeichev et al., 2007). The problem inherent to such single-electrode recordings is that the injected current biases the measurement because of the voltage drop through the electrode. This electrode bias imposes restrictions on the use of advanced electrophysiological techniques like voltage-clamp or dynamic-clamp (Robinson and Kawai, 1993; Sharp et al., 1993; Prinz et al., 2004), which require injection of a current dependent on the simultaneously recorded $V_m$. In the present paper, we introduce a method to compensate for the electrode bias with unprecedented accuracy, based on a model of the electrode interfaced in real time with the electrophysiological setup.

Electrode compensation circuits implemented in intracellular amplifiers usually reflect the assumption that the electrode is a simple RC (resistor and capacitor) circuit (Thomas, 1977), but this simplification does not account for distributed capacitance and the resulting compensation produces artifactual voltage transients (Figure 1A, Bridge compensation, middle). In situations where the injected current depends on the $V_m$, the artifacts are injected back and can be amplified by the control loop, which leads to oscillatory instabilities (Figure 1A, Bridge compensation, right). For this reason, series resistance is usually compensated off-line during voltage-clamp experiments (Borg-Graham et al., 1998; Wehr and Zador, 2003), but this approach severely reduces the effective control of the cell's $V_m$. Another option for voltage-clamp, and the only one for dynamic-clamp when electrode resistance is high, is to use a discontinuous mode, alternatively injecting current and recording the $V_m$ (Brennecke and Lindemann, 1971, 1974a, 1974b; Finkel and Redman, 1984) with a frequency set by the electrode time constant (typically 1.5-3 kHz with sharp electrodes in our experiments in cortical neurons *in vitro*). Unfortunately, the alternation method is valid only when the electrode response is at least two orders of magnitude faster than the recorded phenomena (Finkel and Redman, 1984), because the membrane response must be quasi-linear in the sampling interval. Moreover, recordings in discontinuous modes are very noisy and sampling frequency is limited, which makes the precise recording of fast phenomena like spikes impossible (Figure 1B).

The Active Electrode Compensation (AEC) method we propose here allows the sampling of the $V_m$ during current injection with a frequency only limited by the speed of the computer used for the digital convolution at the core of the technique (Figure 1C). In the following, we describe and validate the method in demanding experimental situations: high resistance sharp microelectrode recordings for both current-clamp with fast current injection and dynamic-clamp protocols. These examples illustrate that with AEC, it is now for the first time possible, in cases when single high resistance electrodes are used, to inject white noise in a cell at a high sampling frequency, and to accurately inject complex conductance stimuli with the goal of precisely analyzing high frequency features of the cell's response like the onset of spikes.

RESULTS

**A method based on a general model of the electrode**

The essential idea behind the AEC method is to represent the electrode by an arbitrarily complex linear circuit, extract the properties of this circuit for each particular recording (Figure 2A), and actively compensate for the effect of the electrode by subtracting the voltage drop through this circuit from the recording (Figure 2B). The classic compensation methods, 'bridge' compensation and capacitance neutralization, are equivalent to model the electrode as a resistor plus a capacitor (RC circuit), but this model proves too simple in many practical situations. We used a more general model of the electrode as an unknown linear circuit. A particular case of such a linear circuit could be two resistors and two capacitors, as hypothesized by Roelfsema et al. (2001). It could also be much more complicated: in fact, our results show that elements of the amplifier (e.g. filters) should be included in the recording circuit.

The voltage across the electrode $U_e$ is modeled as the convolution of the injected current $I_e$ and a kernel $K_e$ which characterizes the electrode:

$$U_e(t) = (K_e * I_e)(t) = \int_0^{+\infty} K_e(s) I_e(t-s)\, ds$$

Thus the voltage across the electrode depends linearly on all past values of the injected current. This formulation encompasses any linear model, e.g. a circuit with a resistor and a capacitor (the kernel $K_e$ is then an exponential function). For digitally sampled signals, the formula reads:

$$U_e(n) = \sum_0^{+\infty} K_e(p) I_e(n-p) \qquad (*)$$

The procedure consists in two passes: 1) measure the electrode kernel, i.e., the values of $K_e(p)$ (Figure 2A, right), 2) inject and record at the same time in continuous mode, with the true $V_m$ recording obtained by subtracting the voltage across the electrode $U_e$ from the raw recording. The compensation involves a digital convolution which is performed in real time by a computer (Figure 2B).

**Measuring electrode properties in the cell**

For small injected currents, the recorded potential can be expressed as the linear convolution $V_r = V_{rest} + K * I$, where $V_{rest}$ is the resting potential and K is the total kernel comprising both the electrode kernel $K_e$ and the membrane kernel $K_m$. We derive K from the recorded response to a known injected current, then we separate the contributions of the electrode and the membrane (Figure 2A; see http://www.di.ens.fr/~brette/HRCORTEX/AEC/AECcode.html for sample code). Electrode properties before and after cell impalement can be quite different (Figure 2C), so it is essential to estimate the electrode kernel in the neuron by means of this separation.

We inject 5-20 s of noisy current consisting of a sequence of independent random current steps (white noise) at sampling resolution (0.1 ms), with amplitude uniformly distributed between -0.5 nA and 0.5 nA in most cases. In principle any current could be used, provided it is small enough to prevent spiking and nonlinear effects (see Supplementary Methods for details about how to set the current intensity), but our choice was not arbitrary: a uniform amplitude distribution makes the best use of the D/A converters in the acquisition board (provided their range is adjusted accordingly), while using a current with minimum autocorrelation enhances the electrode contribution in the recording relatively to the membrane contribution, because the electrode response is at least one order of magnitude faster than the membrane response. The kernel K is then derived mathematically from the autocorrelation of the current and the correlation between the current and the recorded potential (see Methods).

To split the total kernel K into the electrode kernel $K_e$ and the membrane kernel $K_m$, we use two facts: first, the electrode kernel is very short compared to the membrane kernel, so that after a couple of milliseconds (p>50, i.e. 5 ms at 10 kHz) $K_e(p)$ vanishes and $K(p) \approx K_m(p)$; second, if the injected current is small enough the membrane response is mostly linear and can be approximated on short time scales by a decaying exponential $K_m(t) = \frac{R}{\tau} e^{-t/\tau}$ (t = p Δt, where Δt is the sampling step; $R$ and $\tau$ are passive membrane parameters). Thus we estimate the membrane kernel $K_m$ from the tail of the total kernel K (least square fitting to an exponential) and deduce the electrode kernel $K_e$ (see Methods; Figure 2A; see also Supplementary Figure 1).

In fact, the electrode kernel $K_e$ captures not only the characteristics of the electrode but also of the recording device, i.e. the whole circuit between the digital output of the computer and the tip of the electrode, including all circuits in the amplifier (e.g. capacitance neutralization). All measured electrode kernels consisted of 3 phases (Figure 2C): 1) a short phase where the kernel vanishes, 2) a sharp, but non-instantaneous increase for about 0.2 ms, and 3) a decrease. The first phase corresponds to the feedback delay of the acquisition system and always lasts 2 sampling steps (0.2 ms). The second, non-instantaneous phase also appears when the amplifier is plugged into an electronic circuit, named the *model cell*, consisting of a resistor modeling the electrode and a resistor plus a capacitor modeling the neuron membrane (not shown): its non-zero rise-time is likely due to the electronics of the acquisition system rather than to the electrode properties *per se*. The third phase varies between experiments and, when fitted by an exponential, displays a time constant around 0.1 ms (0.11±0.09 ms, n=67 cells for maximal levels of capacitance neutralization by the amplifier). Lowering the level of capacitance neutralization increases this time constant (Figure 2D).

**Estimating the electrode resistance**

As a first test of the method, we assessed the quality of the estimation of the electrode resistance derived from our measurement of the electrode kernel. The electrode resistance $R_e$ is defined as the ratio of the stationary voltage across the electrode over the amplitude of a constant injected current. It equals the integral of the electrode kernel (or the sum in the digital formulation). In 67 cortical neurons *in vitro*, we compared the $R_e$ estimated from the kernel with the $R_e$ estimated by manually adjusting the 'bridge' compensation on the amplifier and found a difference of only 1.4±4.2% (AEC-bridge; n=67 cells).

The linearity of the electrode is an important assumption of the AEC method. It has been described that for sharp electrodes, $R_e$ can change with the amplitude of injected current, presumably due to differences in ion concentrations inside and around the pipette tip (Purves, 1981). We systematically tested for this effect and found that in 9 out of 23 cases, $R_e$ was not significantly correlated with the amplitude of a constant injected current, especially for low $R_e$ values (Supplementary Figure 2): an initial selection of electrodes in the low range of $R_e$ values acceptable for successful impalement of cortical neurons should provide reasonable numbers of linear electrodes. The AEC method provides a fast, quantitative way of checking for linearity at the onset of a recording.

**White noise current injection**

A first application of the AEC method is the possibility to accurately record the response of a neuron to an injection of white noise current sampled at a high frequency (10 kHz with our system). This type of stimulus has been used to characterize neuronal response properties (Bryant and Segundo, 1976). We confirmed (n=18 injections, in 3 cells) that the subthreshold response of neurons to such an injection corresponds to the theoretical prediction based on the passive parameters of the cell (Figure 3A). The recorded $V_m$ distributions closely matched the predicted distributions (Figure 3C,E)

(8.9±9.9% relative error on the standard deviation of the distributions), and power spectra of the response matched the theoretical power spectra up to a frequency of 1 kHz (Figure 3D). In addition, spikes could be recorded with an excellent temporal resolution when a constant depolarizing current was added to the white noise stimulus (Figure 3B). Attempts to inject a white noise current sampled at 10 kHz with DCC at 1-2 kHz switching frequency failed to match the prediction (Figure 3E), presumably due to aliasing effects.

**Dynamic-clamp**

Our initial motivation for developing the AEC method was to improve the quality of dynamic-clamp performed with single high resistance electrodes. Dynamic-clamp (Robinson and Kawai, 1993; Sharp et al., 1993; Prinz et al., 2004) is an electrophysiological technique in which the current injected into the cell is a function of the recorded $V_m$. This loop between current and $V_m$ allows the mimicking of ion channels opening in the membrane. The current flowing through such channels at time *t* depends on both the channels' total conductance $g(t)$ and the driving force $V_m(t) - E_{rev}$ (where $E_{rev}$ is the reversal potential for the considered ions): $I(t) = g(t) \times (V_m(t) - E_{rev})$. In dynamic-clamp experiments, intrinsic or synaptic ion channels are modeled by given $g(t)$ and $E_{rev}$ implemented in a computer, and $I(t)$ is calculated in real time using the $V_m(t)$ of the recorded cell. It is crucial, in this equation, to use the real $V_m$ of the cell, uncontaminated by electrode artifacts which can lead to oscillatory instabilities or, simply, inaccurate results. We tested the performance of the AEC method with three different dynamic-clamp protocols of increasing complexity, comparing to the only alternative method, DCC, and, whenever possible, to theoretical predictions of the response.

*Square conductance pulses*

We first injected a simple conductance stimulus for which the cell's response can be computed analytically, assuming we are in the passive regime and the leak parameters of the cell are known (from the response to small current pulses). This enabled us to compare the responses obtained in AEC and in DCC with theoretical predictions. The stimulus was a square wave of alternating "excitatory" ( $E_{excitation} = V_{rest} + 10$ ) and "inhibitory" ( $E_{inhibition} = V_{rest} - 10$ ) conductance pulses (Figure 4A). Different conductance amplitudes (range 10-100 nS) and frequencies (range 10-1000 Hz) were scanned (n=57 AEC-DCC pairs in total, in 8 cells).

To avoid alignment problems and to separate the quality of the response shapes from the amount of recording noise, we represented the responses in a phase space where they appear as noisy squares and can be compared to theoretical predictions (see Methods, Supplementary Methods and Figure 4B). Three error measures were derived from the comparison of the actual response to the theoretical prediction: the "side" measure quantifies the amplitude of the Vm response, the "tilt" measure quantifies the shape of the response (between a triangular wave and a square wave), and the "distance" measure quantifies the amplitude of the noise around an average response. All those error measures are significantly lower for AEC (tilt error: 5.7±9.2°; side error: 21.2±34%; distance: 0.2 ±0.16 mV) than for DCC (tilt error: 14.6±11.8°; side error: 66.6±86%; distance: 0.43±0.49 mV) ($P<0.0001$ for all three error measures, n=56). Figure 4c (top and middle) shows measured vs. theoretical values for tilt and side. In addition, for all three measures, the advantage of AEC over DCC grows with the waveform's frequency (Figure 4C, bottom; linear regression analyses: $P<0.0001$ for tilt and side error difference, $P=0.0066$ for distance difference), indicating that AEC allows to partly overcome the limitations due to the low sampling frequency in DCC. Low frequency aliasing artifacts in DCC at very high stimulus frequencies are the most striking example of these limitations (Figure 4D).

*Colored conductance noise*

We then injected a colored conductance noise consisting of two stochastic variables, $g_e(t)$ for excitation and $g_i(t)$ for inhibition (Figure 5A), mimicking cortical synaptic background activity as seen *in vivo* (Destexhe et al., 2001) (n=6 AEC-DCC pairs, in 5 cells). Previously derived expressions for the $V_m$ distribution during colored conductance noise injection with known parameters (Rudolph and Destexhe, 2003; Rudolph et al., 2004) and for the $V_m$ power spectrum (Destexhe and Rudolph, 2004) were used to compare responses recorded in both AEC and DCC with a theoretical prediction (see

Supplementary Methods). A very good match was observed between the predicted average $V_m$ and the average $V_m$ measured both in AEC (0.5% average relative error, range 0.003-1.2%) and in DCC (0.5% average relative error, range 0.1-1.8%, no significant difference when compared with AEC, $P$=0.87) (Figure 5B,C left). The measured standard deviations are slightly higher than the prediction (Figure 5B,C right), both in AEC (14.7% average relative error, range 5-21.6%) and in DCC (18.8% average relative error, range 7.7-31%), and the error difference between the two methods, albeit small, is significant ($P$=0.028). The most striking difference between the two methods, however, showed in the frequency content of the $V_m$ fluctuations. In 4 out of 5 cells, the power spectral density (PSD) of the $V_m$ in AEC could be fitted very well with the theoretical template, which provided a good match up to the frequencies where recording noise becomes important (Figure 5D): the PSD scaled in $f^{-4}$ in the high frequencies as predicted. In DCC, the $f^{-4}$ scaling was never observed, the exponent was always smaller (Figure 5E), showing that the correct frequency scaling could only be obtained in AEC (Figure 5F; except for one cell in which all methods yielded erroneous scaling).

*Detailed analysis of spikes*

Finally, we compared spikes recorded in both DCC and AEC (n=7 AEC-DCC pairs, in 4 cells) during a realistic dynamic-clamp protocol (Shu et al., 2003; Wolfart et al., 2005): AMPA inputs (Destexhe et al., 1998) of 5 different amplitudes occurred in a randomized fashion, at 10 Hz, in a synaptic background of fluctuating excitation and inhibition modeled as colored conductance noise (Destexhe et al., 2001). To illustrate the new possibilities offered by high resolution recording of spikes using AEC (Figure 6A, 'AEC'), we performed an analysis of spike threshold variability. In cortical neurons, the $V_m$ value at spike threshold has been shown to correlate with the slope of the preceding depolarization in a way such that faster depolarizations evoke spikes from a more negative threshold (Azouz and Gray, 2000; de Polavieja et al., 2005; Wilent and Contreras, 2005b) . We find such a significant negative correlation in all our AEC recordings (average slope of linear regression: -1.4 ms, range -3.6 to -0.5 ms; average coefficient of regression: 0.407, range 0.137 to 0.621), similar to the example shown in the top row of Figure 6b ('AEC'). This analysis cannot be done with DCC recordings because spike threshold is not detected with enough accuracy (Figure 6A, 'DCC'): the time of spike onset is not locked to the DCC (low) sampling frequency, and thus the first $V_m$ value to be effectively recorded after spike onset is only weakly correlated with the actual $V_m$ value at spike threshold. Instead of detecting the threshold, one detects an approximately random value at the beginning of the spike, and so the slope-"threshold" correlation (average slope of linear regression: 0.5 ms, range 0.07 to 1.43 ms; average coefficient of regression: 0.177, range 0.028 to 0.475) is either non-significant (2/7 injections) or positive (5/7 injections): this relation (Figure 6B, 'DCC') reflects the fact that $V_m$ and $V_m$ slope are positively correlated along the rising phase of a spike. Smoothing the DCC trace can make the spikes look better by eye (Figure 6A, 'sDCC'), but cannot retrieve high frequency information like spike threshold, and so the results of slope-threshold analysis are similar to the ones obtained from a raw DCC trace (significant positive correlation in all 7 cases; average slope of linear regression: 2.8 ms, range 2.1 to 4.1 ms; average coefficient of regression: 0.787, range 0.686 to 0.895) (Figure 6B, 'sDCC').

In addition, we compared spontaneous spikes recorded without any current injection in one cell (Figure 6C, left, red, $V_{spont}$) to spikes evoked in dynamic-clamp in the same cell (Figure 6C, left, blue, $V_{AEC}$; right, black: uncompensated $V_r$ and injected current (bottom)). Spikes recorded using AEC during simultaneous current injection were very similar to spontaneous spikes (compare red and blue), showing that the dramatic influence of the electrode response on spike shape (black, $V_r$) is efficiently compensated by AEC.

**DISCUSSION**

In this paper, we have introduced a new intracellular recording strategy, which we called Active Electrode Compensation, or AEC. We have examined different recording situations of increasing levels of complexity, and in each case, compared AEC recordings with traditional recording techniques, as well as with theoretical predictions when possible. We discuss below the novelty and the domain of applicability of this method.

The AEC method relies on a more powerful model of the electrode than the simple RC model at the basis of most electrode compensation techniques (Purves, 1981; de Sa and MacKay, 2001; Sherman et al., 1999; http://www.moleculardevices.com/ pages/instruments/axon_guide.html; but see Roelfsema et al., 2001), and thus allows accurate intracellular recordings at a high sampling frequency during simultaneous current injection, uncontaminated by capacitive transients. Some new commercial amplifiers propose improved electrode compensation for voltage-clamp, but they do not address the same problem as we do here: the VE-2 patch amplifier (Alembic instruments) implements an adjustable model of the electrode, as we do, and an improved algorithm for controlling the cell's voltage after subtraction of the electrode voltage, but the electrode model remains a simple RC one and the method fails when the electrode's behavior is more complex (Sherman et al., 1999); other amplifiers (SEC, NPI) implement the supercharging technique (Strickholm, 1995; Muller et al., 1999) to speed up the electrode's response by adding brief current pulses to the command current in discontinuous single-electrode voltage-clamp, and thus allow a higher switching frequency, but this technique modifies the total current injected into the cell and so it is not possible to use it in dynamic-clamp to inject a finely controlled current mimicking intrinsic or synaptic ion channels. Besides, all these techniques, which are implemented on manually calibrated analog circuits, are based on simple electrical models of the electrode which cannot fully take the complexity of real electrodes into account. The AEC method uses an arbitrarily complex linear model to compute the electrode's response, does not distort the injected current, does not require manual calibration, and its digital implementation is possible on any computer system adapted to run in real time and interfaced with any standard electrophysiological setup.

The main condition that has to be met for the method to work accurately is the linearity of the electrode and of the whole recording chain between the electrode and the computer (amplifiers and filters), for the range of expected voltages and currents. By estimating electrode properties at different levels of constant injected current, we have confirmed that it is possible to find electrodes displaying no significant non-linearity for a 60-110 M$\Omega$ range of electrode resistances (as measured before impalement). For preparations requiring even higher electrode resistances, it might prove necessary to expand the method in order to account for some non-linearities. As to the rest of the recording chain, its linearity has to be checked in a model electronic cell prior to the use of the method: this condition was satisfied for the Axoclamp 2B and the CyberAmp 380 signal conditioner we tested (not shown).

Other requirements are similar to the requirements of classical methods like 'bridge' compensation or DCC. For optimal separation between cell and electrode kernels, their time constants have to be maximally different, which means that the capacitance neutralization provided by the amplifier should be used optimally like in the other methods. However, numerical simulations show that AEC can work with electrodes only 10 times faster than the membrane, while DCC requires much faster electrodes (Supplementary Figure 3; see also Brette et al., 2007). As with the other methods, if electrode properties change during the recording, the compensation has to be re-adjusted, however a 5s white noise injection was sufficient in our hands to get a reliable electrode kernel estimation and so this re-estimation is not more time-consuming than with 'bridge' compensation or DCC. This minimal length of white noise injection could increase if the cell's response is noisier (*in vivo*), so this parameter should be optimized for each preparation. However, once the right parameters like kernel size and white noise injection length are chosen for a given preparation, the procedure is fully automatic.

Two limitations have to be mentioned, even if they are present in all other recording modes including two-electrode recordings. First, the injected current is still filtered by the electrode. However, in the case of the AEC method, the true injected current can be approximately estimated off-line from the knowledge of the electrode kernel. Second, strong current generated by the cell (i.e. during spikes) affects the electrode response but is not taken into account by the compensation method, and so spikes are still filtered by the electrode (Supplementary Figure 3). This last effect can be reduced by the capacitance neutralization of the amplifier, which is a second reason for using this control together with AEC.

The tests we performed *in vitro* indicate that while for low frequency information, the DCC method presents no severe disadvantage over the high-resolution AEC method, applications requiring high-frequency sampling in single high-resistance electrode recordings are only possible with AEC: we showed that recording responses to white noise current injection at 10 kHz is now feasible, and that dynamic-clamp gains an unprecedented accuracy, allowing to analyze precise features of the spiking response during a finely controlled *in vivo*-like conductance-based stimulus. Moreover, our simulations (Supplementary Figure 3) show that AEC could be used when electrode and membrane time constants are not very different, while DCC cannot be applied in this case.

The extension of accurate dynamic-clamp techniques to new preparations, especially *in vivo*, where either sharp electrodes (Steriade et al., 2001; Crochet et al., 2006; Wilent and Contreras, 2005a; Higley and Contreras, 2007; Haider et al., 2007; Paz et al., 2007) or high series resistance patch electrodes (Borg-Graham et al., 1998; Wehr and Zador, 2003; Mokeichev et al., 2007) have to be used, is only one of the potential applications of the AEC method. We are currently exploring its applicability to single-electrode voltage-clamp protocols. Dendritic and axonic patch-clamp *in vitro*, requiring much finer and thus higher series resistance electrodes than classical whole-cell recordings (>20 MΩ, Davie et al., 2006), could also be a potential field of application. Finally, it should be stressed that the current temporal resolution of the method (10 kHz) is only limited by computer processor speed: it could be increased through implementation on a devoted chip, and its advantage over classical discontinuous methods can be expected to grow substantially in the coming years.

**Acknowledgments:**

This work was supported by CNRS, ANR (HR-CORTEX grant), ACI, HFSP and the European Community (FACETS grant FP6 15879). Z.P. gratefully acknowledges the support of the FRM.

**EXPERIMENTAL PROCEDURES:**

Supplementary Methods provide an extended version with additional details.

**Biological preparation.** We prepared 380 µm-thick coronal or sagittal slices from the lateral portions of 4–12 week old guinea-pig (CPA, Olivet, France) occipital cortex, as well as from adult ferret (Marshall, France) occipital cortex in some early experiments, as described previously (Rudolph et al., 2004; Pospischil et al., 2007). Slices were maintained in an interface style recording chamber at 33–35°C in slice solution containing (in mM) 124 NaCl, 2.5 KCl, 1.2 $MgSO_4$, 1.25 $NaHPO_4$, 2 $CaCl_2$, 26

NaHCO$_3$, and 10 or 25 dextrose, and aerated with 95% O$_2$-5% CO$_2$ to a final pH of 7.4. Intracellular recordings were performed in all cortical layers after 2h of recovery.

**Electrophysiology.** Sharp electrodes for intracellular recordings were made on a Sutter Instruments P-87 micropipette puller from medium-walled glass (WPI, 1BF100) and beveled on a Sutter Instruments beveller (BV-10M). Micropipettes were filled with 1.2–2 M potassium acetate – 4 mM potassium chloride and had resistances of 65–110 MΩ after beveling. An Axoclamp 2B amplifier (Axon Instruments) was used either in continuous current-clamp ('bridge') mode or in discontinuous current-clamp (DCC) mode. In both cases, the capacitance neutralization was set at the maximal possible value to achieve the fastest possible electrode charging time.

**Real-time computer implementation.** We used the hybrid RT-NEURON environment (developed by G. Le Masson, INSERM 358, Université Bordeaux 2), a modified version of NEURON (Hines and Carnevale, 1997) running under the Windows 2000 operating system (Microsoft). To achieve real-time electrode compensation and simulation of synaptic inputs (dynamic-clamp) as well as data transfer to the PC for further analysis, we used a PCI DSP board (Innovative Integration), which constrains calculations made by NEURON and data transfers to be made with a high priority level by the PC processor. The DSP board allows input and output signals to be processed at regular intervals (0.1 ms time resolution). The full RT-NEURON code used in our experiments, as well as some sample code implementing the AEC on different platforms, can be found at http://www.di.ens.fr/~brette/HRCORTEX/AEC/AECcode.html.

**Data analysis.** All values are given as average ± standard deviation or, for small sample size, as average and range. A *P*-value <0.05 was required for statistical significance. Passive neuron parameters (input resistance and membrane capacitance) were derived from responses to small current pulses, and E$_{leak}$=V$_{rest}$.

*White noise injection:* We compared the standard deviations of recorded $V_m$ distributions with theoretical values of standard deviations predicted from the noise parameters and passive neuron parameters (formulas in Supplementary Methods).

*Square conductance pulses:* Recordings were analyzed on phase plots of $V_m(t+T/4)$ vs. $V_m(t)$ (where T is the period of the waveform), where predicted responses for a passive membrane are squares (formulas in Supplementary Methods). We fit an optimal square to the experimental phase plot, then compared the length of the side and the tilt between the optimal square and the theoretical square. The level of noise in the experimental trace was quantified by the average distance (in mV) of the data points to the optimal square. Differences between AEC and DCC were tested for significance with the Wilcoxon non-parametric paired test. Correlations between measures derived from the phase plots and stimulus parameters were evaluated using simple linear regression analysis: the *P*-values given correspond to the null hypothesis that the slope of the linear regression is 0 (two-tailed t-test).

*Colored conductance noise:* Theoretical Vm distributions were computed using a steady-state solution of the passive membrane equation (Rudolph et al., 2004). Differences between AEC and DCC were tested for significance with the Wilcoxon non-parametric paired test. Theoretical templates of the PSD (Destexhe et al., 2004) were fit to the data using a simplex fitting algorithm (Press et al., 1993).

*Spike onset analysis:* Spike threshold was detected using a threshold of 83.3 mV/ms on the derivative of the Vm (similar to Azouz and Gray, 2000). The smoothed DCC trace was obtained with 9-point sliding averaging. The slope of the depolarization preceding the spike was computed by a linear regression on 2.4 ms before the spike. The significance of the correlation between slope of depolarization and spike threshold was assessed using a non-parametric Spearman correlation test.

**Estimation of the electrode kernel.** In order to probe the electrode, we inject a known time-varying current I(n) (n is the number of the sampling step) and measure the response $V_r(n)$ for a duration of N sampling steps (we used 5-20 s, i.e., 50,000-200,000 steps). The kernel K is calculated so that the convolution K*I(n) is the best estimation of $V_r(n)$ in the least square sense. In practice the kernel K is finite and consists of M sampling steps (we used M = 150-200, corresponding to 15-20 ms). For algorithmic reasons (see Supplementary Methods), the injected current I(n) must vanish in the last M sampling steps of the recording. Then the optimal kernel values correspond to the solutions of a matrix problem AX=B, where the coefficients of matrix A and vector B are the autocorrelation coefficients of the current ($\sum_{n=0}^{N} I(n)I(n-p)$) and the current-voltage correlation coefficients ($\sum_{n=0}^{N} V_r(n)I(n-p)$). These coefficients can be calculated recursively online without storing the values $V_r(n)$ and $I(n)$. The matrix A has a special structure (symmetrical Toeplitz matrix), so that the matrix equation can be solved very efficiently with the Levinson algorithm (Press et al., 1993).

The kernel we obtain combines the electrode kernel $K_e$ and the membrane kernel $K_m$. For small currents, the membrane potential can be expressed as the convolution $V_m = V_{rest} + K_m*I_m$, where $I_m$ is the current entering the membrane. The injected current $I_e$ is filtered through the electrode before entering the membrane. We approximate this filtering by the convolution $I_m=(K_e/R_e)*I_e$, where $R_e=\int K_e$ is the electrode resistance. Thus, the total filter K that we measured can be expressed as:

$$K = K_e + K_m * \frac{K_e}{R_e} \qquad (**)$$

In order to retrieve the electrode kernel $K_e$, we need to determine the membrane kernel $K_m$ and invert the relationship above. We approximate $K_m$ by an exponential function: $K_m(t)=\frac{R}{\tau}e^{-t/\tau}$, and we estimate the time constant τ by fitting an exponential function to the tail of the kernel K (typically t > 5 ms), which is correct if the electrode is faster than the membrane. From this fit, we obtain an estimation R' of the membrane resistance. It is an overestimation because the electrode delays the response of the membrane. The electrode resistance $R_e$ is then estimated as $R_e'=\int K - R'$. In practice, we have only a truncated version of the full kernel (typically the first 15-20 ms), so that only part of the membrane resistance must be subtracted. Once estimates have been derived for R, $R_e$ and τ, it is possible to deduce $K_e$ from K. We solve equation (**) by applying the Z-transform, which transforms convolutions into multiplications (see Supplementary Methods).

Finally, we refine the electrode kernel as follows. If the estimates of R and $R_e$ were correct, then the tail of the kernel (t > 5 ms) should vanish. If there is a positive remainder, then R was underestimated; if there is a negative remainder, then it was overestimated. Therefore, in order to reach the best precision, we reiterate the procedure with different estimates for R (and corresponding estimates for $R_e$) using the golden section search algorithm (Press et al., 1993), so as to minimize the tail of the estimated kernel $K_e$.

**FIGURE CAPTIONS:**

**Figure 1.** Compensating for the electrode response during simultaneous current injection and recording of neuronal membrane potential ($V_m$).

**(A)** 'Bridge' compensation performed by the amplifier (left). The capacitive properties of the electrode lead to a capacitive transient at the onset of the response to a current step (middle; scale bars: 5 nA, 10 mV, 10 ms). A loop is established between $V_m$ recording and current injection when inserting virtual conductances G in dynamic-clamp (right). When fast fluctuating conductances are inserted, the transients lead to a strong 'ringing' oscillation in the recorded potential $V_{Bridge}$ and the injected current I (scale bars: 100 mV, 20 nS, 4 nA, 100 ms). With square conductance pulses, a dampened oscillation can be seen (scale bars: 5 mV, 10 nS, 0.2 nA, 5 ms).

**(B)** Discontinuous current clamp (DCC; see text for details). There is no capacitive transient at the onset of the response to a current step (middle; scale bars: 5 nA, 10 mV, 10 ms). Conductance injection using dynamic-clamp can be performed (right; scale bars: 10 mV, 20 nS, 5 nA, 100 ms) without oscillations, but the sampling resolution of the $V_m$ is low, as seen when zooming in on single spikes (inset; scale bars: 10 mV, 2 ms).

**(C)** Active electrode compensation (AEC), a new method for high-resolution $V_m$ recording during simultaneous current injection. This digital compensation is performed in real-time by a computer (left). No capacitive transient is seen at the onset of the response to a current step (middle; scale bars: 5 nA, 10 mV, 10 ms). Conductance injection using dynamic-clamp (right; scale bars: 10 mV, 20 nS, 5 nA, 100 ms) is performed with a high $V_m$ sampling frequency (10 kHz), so that the shape of single spikes can be resolved (inset; scale bars: 10 mV, 2 ms).

**Figure 2.** The two stages of the AEC method.

**(A)** Electrode properties are estimated first: white noise current (scale bar: 0.5 nA) is injected into the neuron, and the total response $V_r$, corresponding to the sum of the membrane potential $V_m$ and the voltage drop across the electrode $U_e$, is recorded (left; scale bars: 10 mV, 10 ms). The cross-correlation between the input current and the output voltage gives the kernel (or impulse response) of the neuronal membrane + electrode system (full kernel, right). This full kernel is separated into the electrode kernel and the membrane kernel.

**(B)** The electrode kernel is then used in real time for electrode compensation: the injected current (scale bar: 5 nA) is convolved with the electrode kernel to provide the electrode response $U_e$ to this current. $U_e$ is then subtracted from the total recorded voltage $V_r$ (scale bars: 100 mV) to yield the $V_m$ ($V_{AEC}$; scale bars: 10 mV, 100 ms).

**(C)** Kernel of the same electrode estimated in the slice before the impalement of a neuron (black), and after impalement, in a cell (red).

**(D)** Electrode kernels obtained in the slice for different levels of capacitance neutralization (the sharpest kernel corresponds to the highest level of capacitance neutralization).

**Figure 3.** White noise current injection using AEC.

**(A)** Example $V_m$ response (scale bars: horizontal left, 10 ms; horizontal right, 1 ms; vertical 5 mV) of a neuron recorded using AEC (blue) and $V_m$ response obtained by simulating the same noise injection (scale bar: 1 nA) in a point model neuron using the leak parameters of the recorded cell (red). The injected current is displayed below (black).

**(B)** Spiking response (scale bars: 50 mV, 100 ms) to supra-threshold white noise current injection. Black ('Bridge'): reconstructed $V_m$ response after bridge balancing, i.e. instead of using the electrode kernel for compensation, the electrode response to the current is modeled as *electrode resistance× injected current*. Blue: $V_m$ response recorded using AEC. Bottom: zoom on spikes recorded using AEC (scale bars: 20 mV, 20 ms).

**(C)** $V_m$ distribution recorded using AEC (blue) and theoretical distribution in response to the same injected current (red).

**(D)** Power spectral density (PSD) of the $V_m$ recorded using AEC (blue) and theoretical PSD in response to the same injected current (red). The PSD of the baseline $V_m$ (gray) shows that the bump in the higher frequencies is not due to AEC, but rather to the power of recording noise reaching the level of the signal.

**(E)** Pooled data: standard deviation of the $V_m$ distributions obtained using AEC (blue, open) or DCC (black, filled), vs. theoretical standard deviation based on leak parameters of the recorded neuron (error bars represent the range of theoretical standard deviations obtained for different estimates of passive cell parameters; solid line: y=x).

**Figure 4.** Conductance square wave injection in dynamic-clamp.

**(A)** Example $V_m$ responses (scale bars: 5 mV, 5 ms) to a square wave (50 nS amplitude) of alternating excitatory (green) and inhibitory (orange) conductances, using AEC (top, blue) or DCC (black, middle). The response obtained by simulating the same conductance injection in a point model neuron using the passive parameters of the recorded cell is shown in red.

**(B)** Phase plots of the $V_m$ responses shown in (a): instead of plotting $V_m$ vs. time, here $V_m$ at time t+T/4 is plotted against $V_m$ at time t (T: period of the injected wave). Red: theoretical phase plot calculated using the stimulus parameters and the cell's leak parameters. Yellow: square providing the best fit to the experimental phase plot. Each square is characterized by its side length S and its tilt relative to the vertical, θ (see Methods).

**(C)** Pooled data from all cells and injection parameters used (AEC: blue open circles; DCC: black filled diamonds): tilt of the recorded phase plot vs. the theoretical tilt (top); side length of the recorded phase plot vs. theoretical side (middle); and difference (DCC – AEC) between tilt and side errors (relative to the theoretical prediction) in AEC and in DCC vs. square wave frequency.

**(D)** $V_m$ response (scale bars: horizontal left, 1s; horizontal right, 2 ms; vertical, 5 mV) to high frequency (500 Hz) conductance wave injection (50 nS amplitude), in AEC (blue, top) and DCC (black, bottom).

**Figure 5.** Colored conductance noise injection in dynamic-clamp.

**(A)** Example $V_m$ responses (scale bars: 5 mV, 100 ms) to fluctuating excitatory (green) and inhibitory (orange) conductances (scale bar: 20 nS), using AEC (blue, left) or DCC (black, right). The response obtained by simulating the same conductance injection in a point model neuron using the passive parameters of the recorded cell is shown in red.

**(B)** $V_m$ distribution using AEC (blue), DCC (black) and theoretical distribution in response to the same injected conductance noise (red).

**(C)** Pooled data: mean of the $V_m$ distributions obtained using AEC (blue, open) or DCC (black, filled), vs. theoretical mean based on leak parameters of the recorded neuron (left); standard deviation of the $V_m$ distributions obtained using AEC (blue, open) or DCC (black, filled), vs. theoretical standard deviation based on leak parameters of the recorded neuron (right). (Error bars represent the range of theoretical values obtained when varying estimates of leak parameters by ±10%; solid lines: y=x).

**(D)** Power spectral density (PSD) of the $V_m$ recorded using AEC (blue) and best fit with the theoretical template for the PSD (red).

**(E)** PSD of the $V_m$ recorded using DCC (black) and best fit with the theoretical template for the PSD (red).

**(F)** Pooled data: root-mean-square (RMS) error of the best fit to the experimental $V_m$ PSDs obtained when using the theoretical template (AEC: blue, open; DCC: black, filled; each point is an average for PSDs obtained from fragments of conductance injections done in the same cell, error bars are standard deviations).

**Figure 6.** Spikes evoked by a complex dynamic-clamp conductance injection.

**(A)** Examples of single spikes recorded in AEC (blue, top) or DCC (black, middle), for different slopes of depolarization preceding the spike. Spikes obtained when the DCC recording is smoothed are shown at the bottom ('sDCC'). Insets: zoom on the onset of spikes (scale bars: 10 mV, 1 ms).

**(B)** Spike threshold vs. slope of depolarization preceding spikes, from injections corresponding to the examples shown in (a). Lines are linear regressions to the clouds of points.

**(C)** Spontaneous spikes (left, red, $V_{spont}$) are compared to spikes evoked in the same cell during dynamic-clamp injection of a fluctuating inhibitory conductance, using AEC (left, blue, $V_{AEC}$; scale bars: 20 mV, 2 ms; dashed line: 0 mV). Fluctuating current flowing through the electrode (right, black, bottom; scale bar: 1 nA; dashed line: 0 nA) affects the total uncompensated response $V_r$ (right, black, top) due to the electrode response, but not the compensated $V_{AEC}$ during the spike (left, blue).

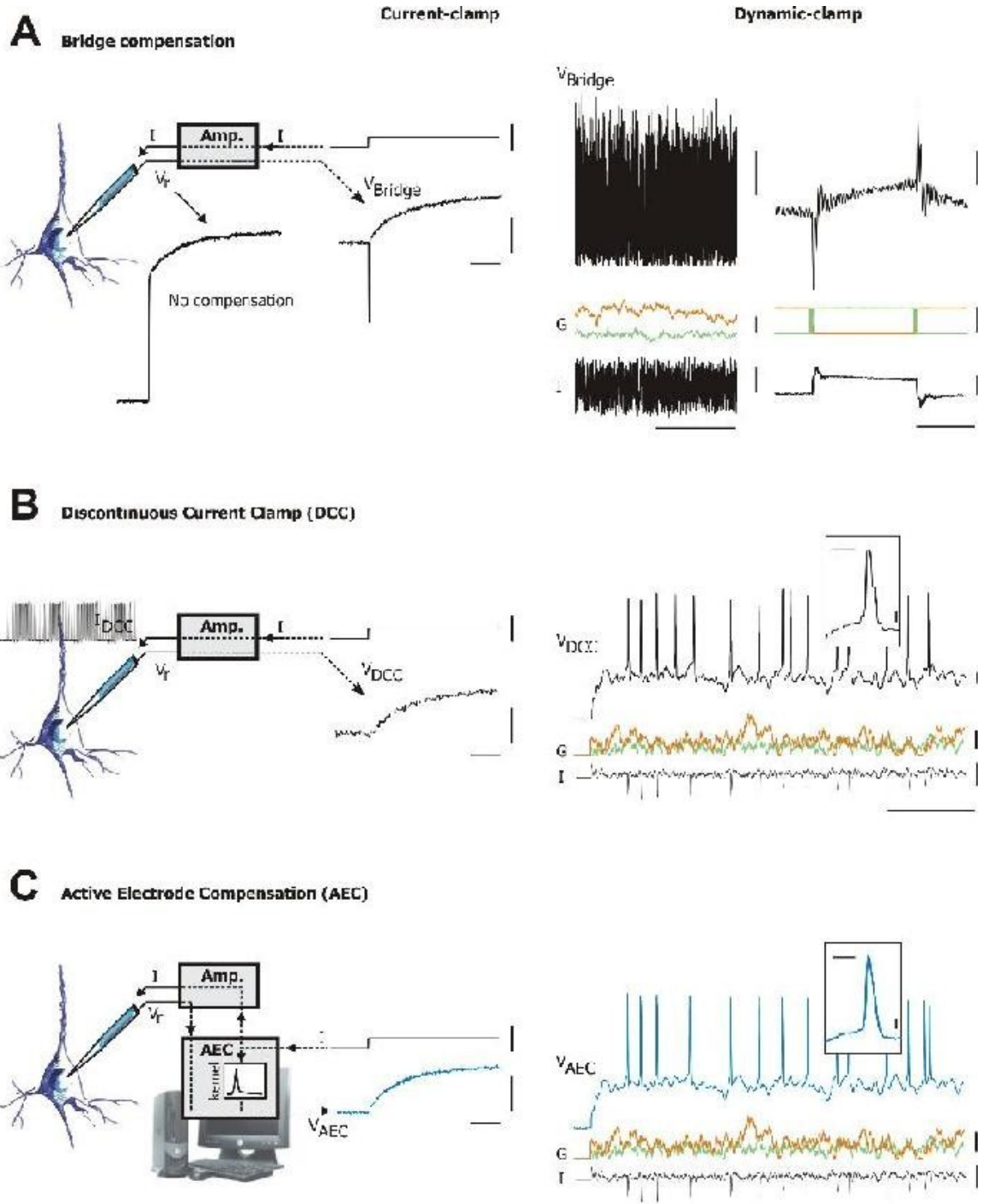

Fig. 1

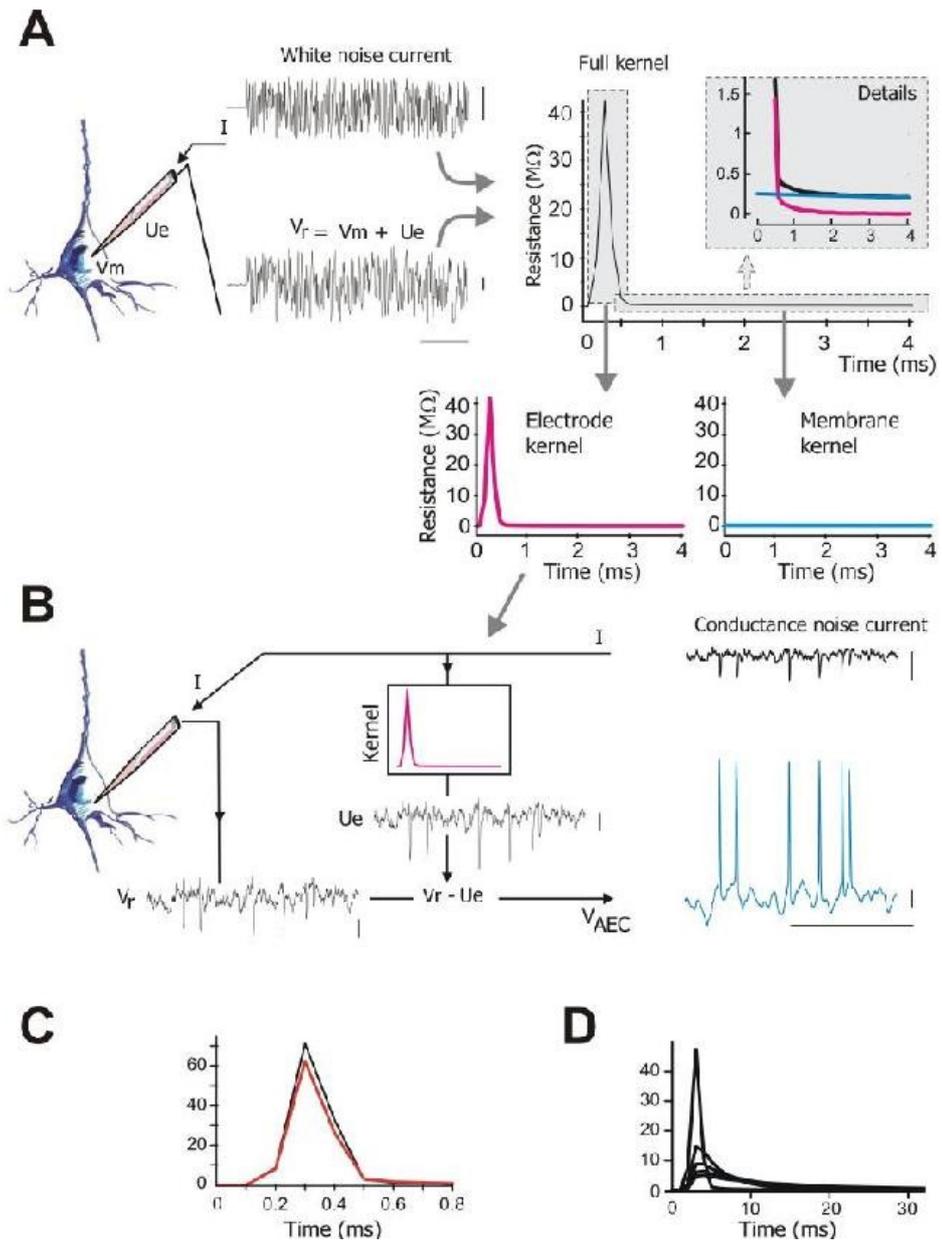

Fig. 2

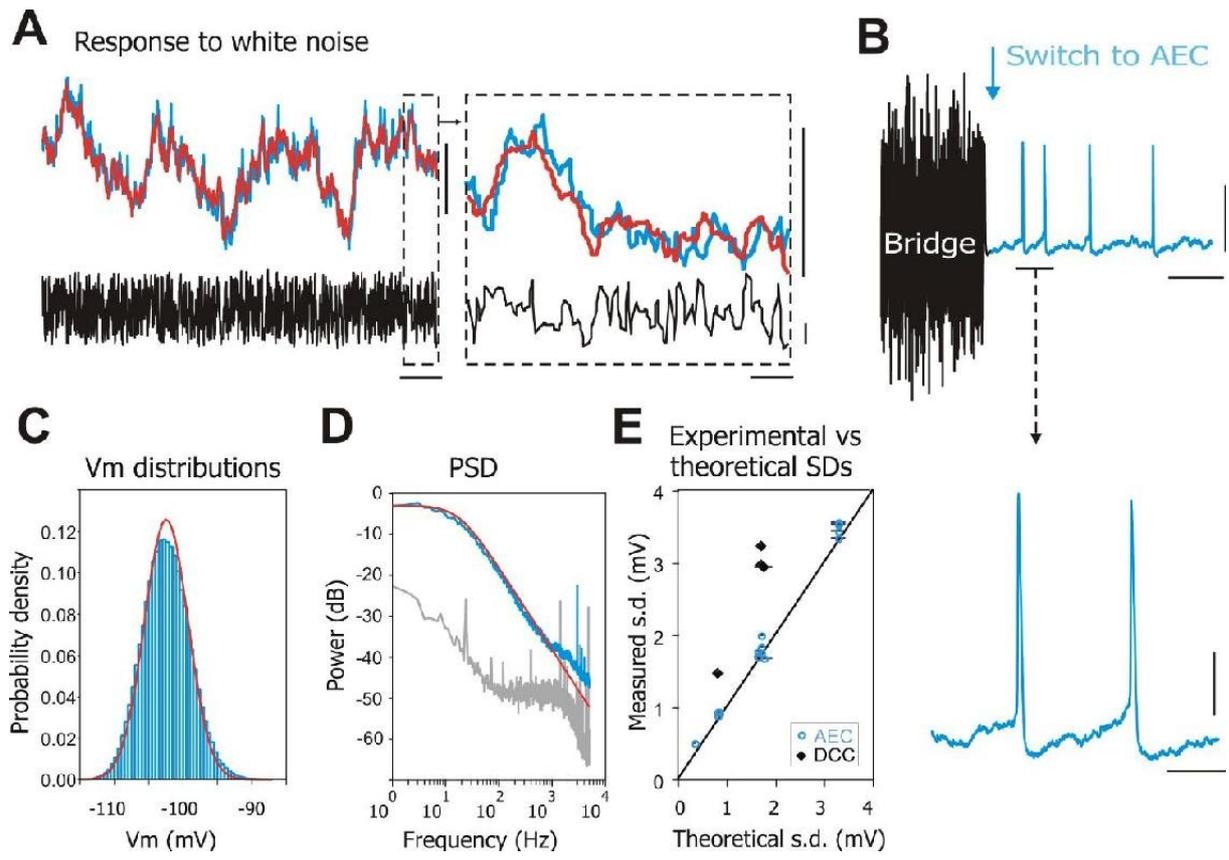

Fig. 3

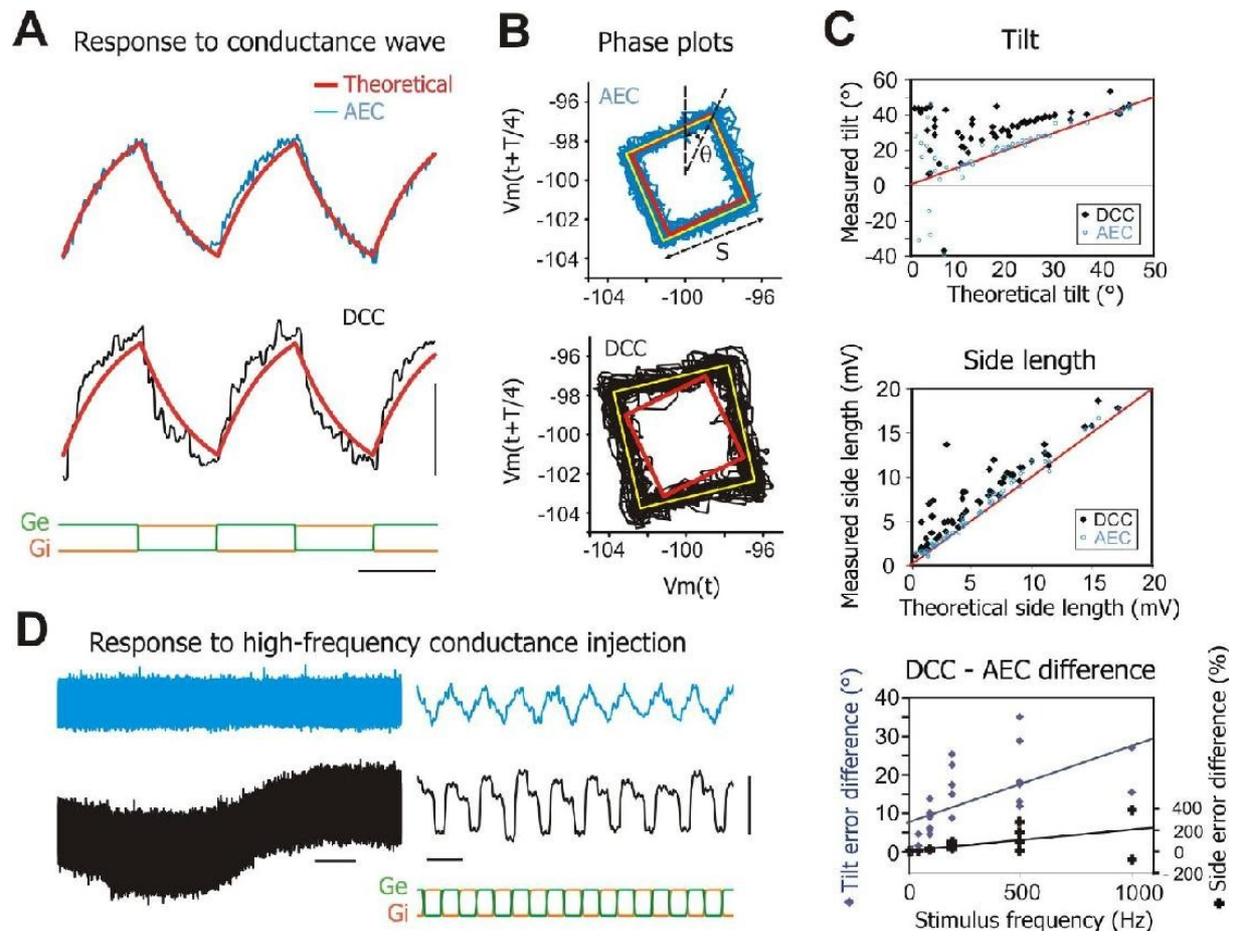

Fig. 4

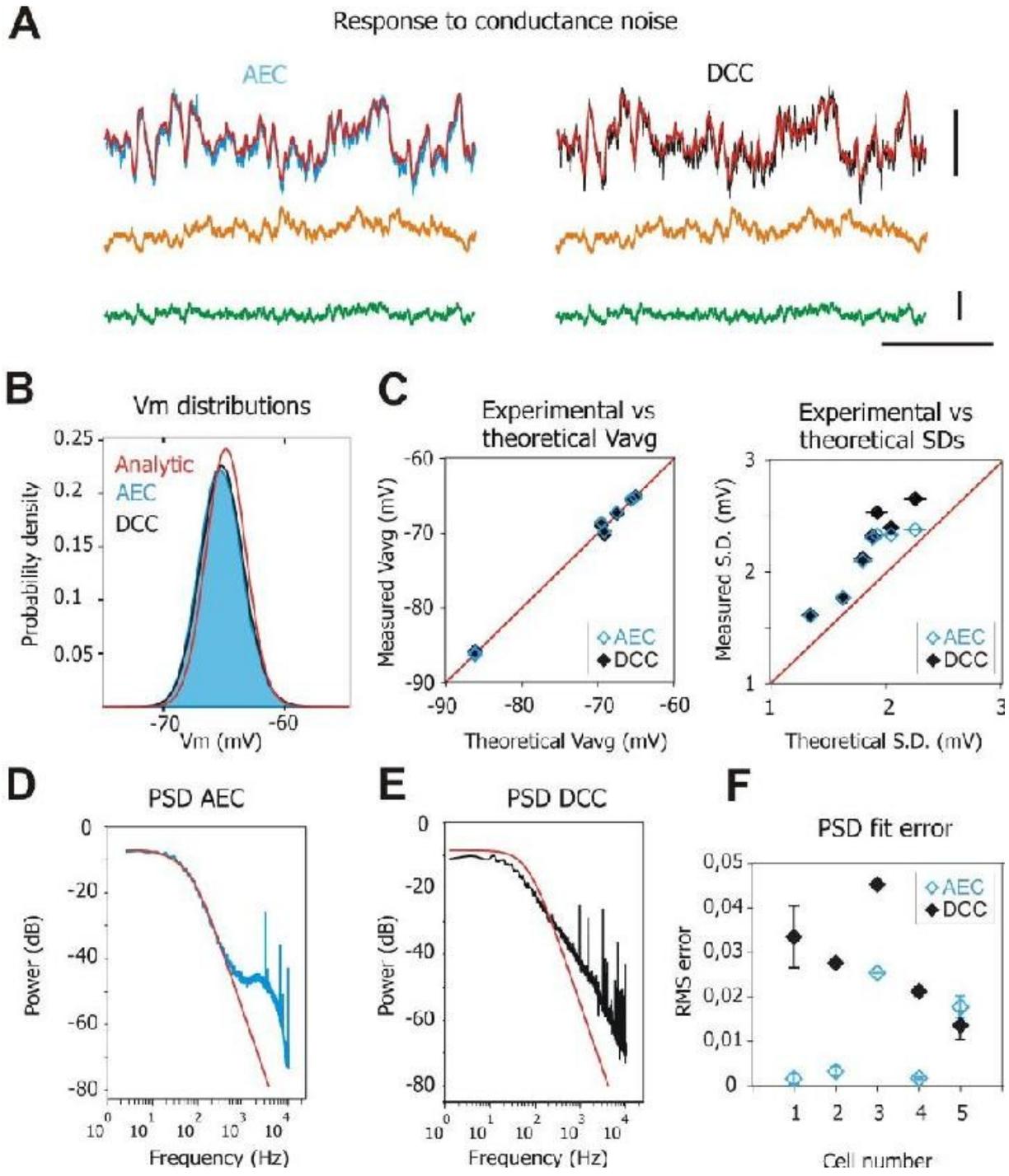

Fig. 5

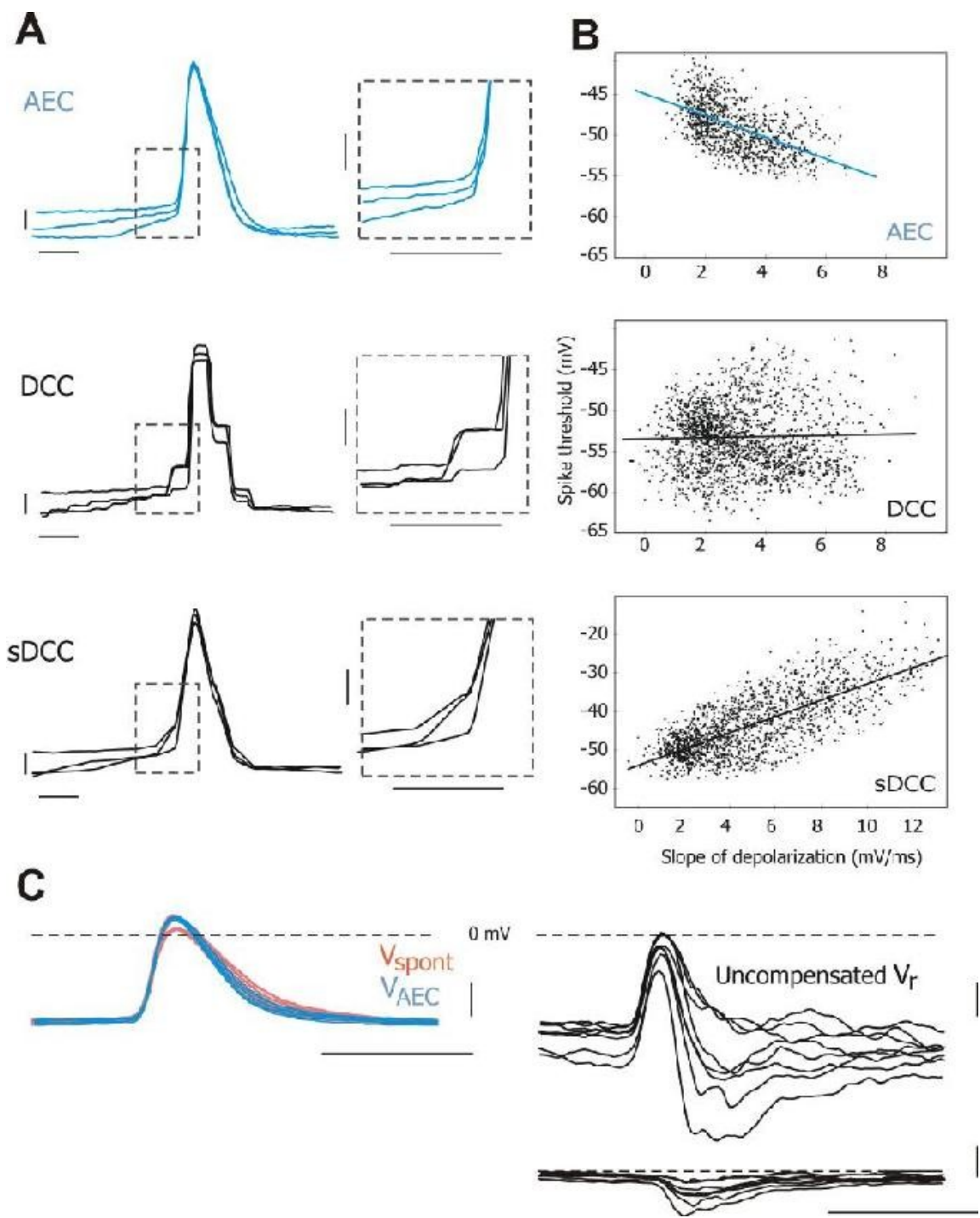

Fig. 6